\documentstyle[12pt]{article}

\def\bi{\bigskip}
\def\noi{\noindent}
\begin{document}

\begin{center}
{\Large \bf SPIN 3/2 INTERACTING FIELDS AND HEAVY BARYON CHIRAL PERTURBATION THEORY\footnote{Work supported by CONACyT under contract 4918-E}}\\[1.5cm]
\end{center}
\vspace{1cm}
\begin{center}
{{\large \bf M. Napsuciale and J.L. Lucio M.}\\[1cm]
\vspace{.8cm}
{\it Instituto de F\'\i sica, Univiversidad de Guanajuato}\\
{\it Apartado Postal E-143, Le\'on, Gto., M\'exico}}
\end{center}
\noindent

\vspace{2cm}

\begin{center}
{\bf Abstract.} 
\end{center}

We ~analyze ~the ~consistency ~of ~the ~Chiral ~Lagrangian ~approach ~to ~the 
~des\-crip\-tion of the spin 3/2 interacting theory. We argue that to lowest 
order in the 1/m expansion, the formalism leads to the appropriated 
constraints and the theory is free of the so called ``off shell" ambiguities.

\setlength{\baselineskip}{1\baselineskip}

\newpage

Chiral Perturbation Theory is a useful tool in the ~description ~of ~low 
~momentum
process involving Goldstone Bosons [1]. When the formalism is extended to
include interactions with Baryons it suffers from two unpleasant 
characteristics i) There is no correspondence between the loop and momentum 
expansion. ii) The expansion parameter turn out to be of order one. A new 
formalism to circumvent the problems arising in Baryon Chiral Perturbation
Theory (BChPT) was ~introduced ~by ~Jenkins ~and ~Manohar (J-M) [2]. The 
approach
of J-M is based on the technics originally developed to describe heavy quark
systems [3], which amounts to consider the baryons as heavy static sources of 
Golstone Bosons. The effective theory is written in terms of definite velocity
fields for which the Dirac equation correspond to a massless baryon. The heavy
baryon Lagrangian has a 1/m expansion, the 1/m effects in the original theory 
are reproduced in the effective theory by including higher dimension operators
suppressed by inverse powers of m. 

\bi

Besides the baryon octet the decuplet is also included in the effective 
theory. In this case the theory involves spin 3/2 fields whose effective mass 
is $\Delta m =M-m$ (m-M masses of the baryon octet and decuplet respectively).
It is possible to construct an effective theory with only octet baryons and  
Goldstone Bosons by integrating out the decuplet fields. Decuplet effects in
the original theory are ~reproduced in the new theory by higher dimension
~operators ~involving ~only ~octet ~baryons ~and ~mesons. ~These ~higher 
~dimension 
operators are suppressed by inverse powers of $\Delta m$. In the real world
the decuplet-octet baryon mass difference $\Delta m \approx 300 MeV$ which 
is small compared to the hadronic scale of 1GeV. The smallness of $\Delta m$
produces an enhancement of the decuplet effects and for that reason it is best
to retain explicit decuplet fields in the effective theory, rather than 
integrate them out. This formalism has been applied to study the $T \to BP$
and $T \to B \gamma$ decays [4] (T,B,P stand for decuplet, baryon and meson
octets respectively). The next to leading order of Baryon Chiral Perturbation
Theory with $S= 1/2$ baryons and Goldstone bosons have also been examined [5].

\bi

It is well known that Quantum Field Theory for spin 3/2 interacting fields 
suffers of serious incosistencies. In particular, within the Rarita-Schwinger 
formalism [6] the R-S spinor $\psi_\mu$ has more degrees of freedom than 
required, which makes the classical Lagrangian non-unique. In fact there exist
a whole family of one parameter Lagrangians ${\cal L} (A)$ from which the   
Dirac-Fierz-Pauli equation and the necessary free field constraints for 
$\psi_\mu$ are obtained [7]. Furthermore, these Lagrangians remain invariant
under point transformations (R-invariance) which mix the spurious spin 1/2
fields contained in $\psi_\mu$. This invariance ensures that the unphysical 
spin 1/2 degrees of freedom have no observable effects. Generalizations of 
this scheme to describe interacting spin 3/2 fields requires that the 
co\-rres\-pon\-ding Lagrangian preserves R invariance, this fact introduces 
new ambiguities into the theory (the so-called ``off-shell" parameters) [8]. It
should also be emphasized that the constraints equations, which ensure the
elimination of the unphysical spin 1/2 degrees of freedom in the free case, 
are spoiled by the interaction.

\bi

It is our purpose in this paper to analyze the consistency of the Chiral 
Lagrangian approach to the description of the spin 3/2 interacting theory. We 
construct an R invariant Lagrangian and argue that only to lowest order in 
the 1/m expansion, the approach leads to the appropriated constraints and the 
theory is free of the so called ``off-shell" ambiguities.

\bi
\bi

\noi {\bf 1 ~~FREE FIELD THEORY}

\bi

Within the Rarita-Schwinger approach, the S=3/2 field is described by a 
spinor-vector $\psi_\mu$ which is obtained from the tensor product of a spinor
and a four-vector. Clearly $\psi_\mu$ involves more degrees of freedom than
required to describe a spin 3/2 field. This redundancy is reflected in the
formalism in the need of the subsidiary conditions:

\begin{eqnarray}
\gamma^\mu \psi_\mu (x) &=& 0 \nonumber \\
\partial^\mu \psi_\mu (x) &=& 0 
\end{eqnarray}

\bi

\noi It has been shown that for the spin 3/2 field, there exist a whole family
of one parameter Lagrangians from wich the equation of motion and the
subsidiary conditions can be derived [7]

\bi

\begin{equation}
{\cal L} (A) = \psi^\mu (x) \{i \partial_\alpha \Gamma^\alpha\,_{\mu\nu} (A) -
M B_{\mu\nu} (A) \} \psi^\nu (x)
\end{equation}

\noi where

\begin{eqnarray}
\Gamma^\alpha_{\mu\nu}(A) &=& g_{\mu\nu} \gamma^\alpha + B \gamma_\mu 
\gamma^\alpha \gamma_\nu + A (\gamma_\mu g^\alpha\,_\nu + g_\mu\,^\alpha
\gamma_\nu ) . \nonumber \\
B_{\mu\nu} (A) &=& g_{\mu\nu} - C \gamma_\mu \gamma_\nu \nonumber \\
A \not = -1/2 &,& B= \frac{3}{2} A^2 +A+\frac{1}{2} , ~~C= 3A^2+3A+1 \nonumber
\end{eqnarray}

\bi

\noi For A=-1/3, ${\cal L} (A)$ reduces to the Lagrangian originally proposed 
by R-S [6]. From (2) the spin $\frac{3}{2}$ field propagator is obtained

\bi

\begin{equation}
i \Delta_{\mu\nu} (P , A)= \frac{\Sigma_{\mu\nu} (P,A)}{P^2-M^2+ i \epsilon}
\end{equation}

\noi 
with
\begin{eqnarray}
\Sigma_{\mu\nu} (P,A)=2M S_{\mu\nu} &-& \frac{1}{6} ~\frac{A+1}{2A+1} 
~\frac{P^2- M^2}{M} \{ \gamma_\mu \bigg(\frac{2P}{M}-\gamma \bigg)_\nu 
\nonumber \\
& + &\bigg(\frac{2P}{M}- \gamma \bigg)_\nu \gamma_\mu 
- \frac{A+1}{2A+1} \bigg( \gamma_\mu \frac{\not P}{M}\gamma_\nu -2\gamma_\mu
\gamma_\nu \bigg) \} \nonumber
\end{eqnarray}

\noi and
$$S_{\mu\nu}= \{- g_{\mu\nu}+ \frac{1}{3} \gamma_\mu \gamma_\nu - \frac{1}{3M}
(\gamma_\mu P_\nu- P_\mu \gamma_\nu)+\frac{2}{3M^2} P_\mu P_\nu \} \frac{\not 
P +M}{2M}$$

\bi

\noi The following properties of the Lagrangian are worth-remarking

\bi
\bi

\noi {\bf -R Invariance.} By construction the Lagrangian (2) is invariant 
under the point transformations:

\bi

\begin{equation}
\psi_\mu \to \psi^\prime_\mu = R_{\mu\alpha} (a) \psi^\alpha \qquad\qquad
A \to A^\prime = \frac{A-2a}{1+4a}
\end{equation}

\bi

\noi where

\bi

$$R_{\mu\nu} (a) = g_{\mu\nu} + a \gamma_\mu \gamma_\nu \qquad\qquad 
a \not = - \frac{1}{4}$$

\bi

The R operator acts only on the spin 1/2 content of $\psi_\mu$. The 
arbitrariness of the spin 1/2 components of $\psi_\mu$ is at the origin
of the family of one parameter Lagrangians (2).

\bi

\noi Useful properties of the R operator are

\begin{eqnarray}
R_{\mu\nu} (a) R^\nu\,_\lambda (b) &=& R_{\mu\lambda} (a+b+4ab); \nonumber \\
R^{-1}_{\mu\nu} (a) &=& R_{\mu\nu} \bigg(- \frac{a}{4a+1} \bigg) , \\
R_{\mu\nu} (0) &=& g_{\mu\nu} ; \nonumber
\end{eqnarray}

\bi
\bi

\noi {\bf - A Independence.} The generalization of (2) to include interactions
should lead to physical quantities which are A independent. This is assured by
the Kamefuchi-O'Raifeartaigh-Salam [9] theorem. A proof of the theorem in this
particular case is given by Urias [7] based on the observation that.

\begin{equation}
{\cal L} (A) = \bar\psi^\mu K_{\mu\nu} (A) \psi^\nu
\end{equation}

\noi where
$$K_{\mu\nu} (A)= R_{\mu\alpha} \bigg( \frac{1}{2} (1+3A)\bigg) {\cal K}^{\alpha\beta} R_{\beta\nu} \bigg( \frac{1}{2} (1+3A) \bigg)$$

\bi

\noi and ${\cal K}^{\alpha\beta}$ is the operator corresponding to
A=- 1/3
$${\cal K}^{\alpha\beta} = (\not P -M) g^{\alpha\beta} + \frac{1}{3} 
(\gamma^\alpha \not P \gamma^\beta - P^\alpha \gamma^\beta - \gamma^\alpha 
P^\beta + M \gamma^\alpha \gamma^\beta ).$$

An application of this factorization property to describe the $\pi p$ elastic 
scattering in the $\Delta$ region has been reported in [10].

\bi
\bi

\noi {\bf 2 ~~INTERACTIONS.}

\bi

We restrict our attention to the strong interactions of spin 3/2 baryons, 
electromagnetic interactions can be introduced by gauging the theory. There 
exist abundant literature [8,12,13], about the inconsistencies of the first 
and second quantized version of the interacting spin 3/2 field theory. Further 
ambiguities associated to the so-called ``off-shell" parameters (to be 
discussed below) have also ~been ~analyzed ~[13]. ~In ~spite ~of ~the 
~difficulties 
previously mentioned, chiral symmetry has been traditionally taken as the 
guiding principle to construct phenomenological Lagrangians aiming to describe
the existing data [11,14]. Further requirements for any sensible Lagrangians 
are:

\bi

\noi {\bf R-Invariance.} The full Lagrangian, ${\cal L} = {\cal L}_{free} +
{\cal L}_{int}$ must be invariant under contact R transformations. This
ensures, trough the K-O-S theorem [9] the A independence of physical
quantities.

\bi

\noi {\bf Constraints.} Once the interactions are included, the Lagrangian 
must lead to the appropiated constraints so that only the correct number
of degrees of freedom to describe a spin 3/2 field are left.

\bi

A convenient procedure to construct R invariant Lagrangian describing 
interacting spin 3/2 fields can be envisaged paralleling the factorization
property (6) of the free Lagrangian. For definitiveness consider the 
$\frac{3}{2}^+ \frac{3}{2}^+ 0^-$ interaction. Using the nonlinear realization 
of chiral symmetry the Lagrangians has the Lorentz structure (In the 
following we omit the SU(3) structure of the interaction as it is of no
relevance for our argument)
$${\cal L}_{\frac{3}{2}\frac{3}{2}0} = \bar \psi^\mu O_{\mu\nu\alpha} \psi^\nu
\Delta^\alpha$$

\bi

Where $\psi_\mu , \psi$ stand for the decuplet tensor and octet baryon matrix
respectively and $\Delta^\nu$ stand for the axial vector field

$$\Delta^\mu = \frac{i}{2} (\xi \partial^\mu \xi^+ - \xi^+ \partial^\mu \xi)$$

\noi with $\xi = exp \bigg( \frac{i}{f_\pi} \pi \bigg) ,\pi$ being the 
pseudoscalar field matrix.

\bi

\noi The simplest possibility for $O_{\mu\nu\alpha}$ is

$$O^S_{\mu\nu\alpha} = g_{\mu\nu} \gamma_\alpha \gamma_{_5}$$

\bi

Since all the Lagrangian related by an R transformation are equivalent [7], a 
family of A dependent Lagrangians is obtained by applying the point 
transformations (5)

\begin{equation}
O_{\mu\nu\alpha} (A) = R_{\mu\beta} (h(A)) g^\beta\,_\sigma \gamma_\alpha
\gamma_5 R^\sigma\,_\nu ( \ell (A))
\end{equation}

\bi

\noi {\it h(A)},$\ell (A)$ ~are, ~in ~principle, ~arbitrary ~functions ~of ~A, 
~however ~the ~requirement of R invariance of the Lagrangian limits $h(A)$ and 
$\ell (A)$ to the class of functions for which the following identity holds:

\begin{equation}
R_\mu\,^\delta (a) O_{\delta\beta\alpha} (A^\prime) R^\beta\,_\nu (a) = O_{\mu\nu\alpha} (A)
\end{equation}

\bi

\noi Using the composition properties (5) of the R operators, (8) reduces to 
the  pair of equations   

$$a+(1+4a)h(A^\prime)= h(A) ; ~~a+(1+4a)\ell (A^\prime)= \ell (A)$$

\bi

For linear functions of A, the solution for $\ell(A), h(A)$ are (X,Y arbitrary
parameters); 

\begin{equation}
\ell (A) = \frac{1}{2} (1+ 4X) A+ X; ~~h (A) = \frac{1}{2} (1+4Y) A+Y
\end{equation}

\bi

A ~similar ~analysis ~can ~be ~applied ~to ~the ~$\frac{3}{2}^+ \frac{1}{2}^+ 
0^-$ ~interactions ~with ~similar ~results ~[8], the R-invariant vertex 
depending on
A and a new arbitrary parameter Z. Therefore, the R invariant Lagrangian 
involves, besides A, the so-called ``off-shell" parameters X,Y,Z.

\bi

There ~exist ~controversy ~in ~the ~literature ~regarding ~the ~quantization 
~of ~spin ~3/2 ~interacting ~theory. ~Whereas ~some ~authors ~[8] ~claim ~that
~the ~constraints and the ~requirement of a ~consistent ~quantization ~pick ~up
particular values for ~the ``off-shell" ~parameters a carefull ~analysis [13] 
~shows that ~negative ~norm states cannot be ~avoided. ~Alternative 
~approaches [14], disregard the arbitrariness of the spin 1/2 componets of 
$\psi_\mu$, take X,Y,Z as free ~parameters and use ~them to fit the 
~experimental ~data.

\newpage

\noi {\bf 3 ~~HEAVY BARYON CHIRAL PER\-TUR\-BA\-TION THEO\-RY (HBChPT).}

\bi

We have seen that Baryon Chiral Perturbation Theory necessarily involves
interacting spin 3/2 fields, and we have also summarized the problem faced by 
the interacting spin 3/2 field theory. The next step is to analyze the new 
ingredients that the heavy fermion approach brings to the scene. According to 
the results of the previous section, to lowest order in the chiral expansion,
the Lagrangian is

\begin{equation}
{\cal L} = {\cal L}_{free} + {\cal L}_{\frac{3}{2}\frac{1}{2}0} + {\cal L}_{\frac{3}{2}\frac{3}{2}0}  
\end{equation}

\bi

\noi the free Lagrangian is given in (2) and the R-invariant interactions read

\begin{eqnarray}
{\cal L}_{\frac{3}{2}\frac{1}{2}0} &=& i {\cal C} \bar\psi^\mu O_{\mu\nu} 
(A,Z)\psi \Delta^\nu + h.c. \nonumber \\
{\cal L}_{\frac{3}{2}\frac{3}{2}0}&=&i{\cal H} \bar\psi^\mu O_{\mu\nu\alpha}
(A,X,Y) \psi^\nu \Delta^\alpha + h.c. \nonumber
\end{eqnarray}

\bi

where ${\cal C}$ and ${\cal H}$ are coupling constants. $O_{\mu\nu\alpha}
(A,X,Y)$ is given by:

$$O_{\mu\nu\alpha} (A,X,Y) = R_{\mu\beta} (f(A,X)) g^\beta\,_\sigma \gamma_\alpha \gamma_5 R^\sigma\,_\nu (f(A,Y))$$

\bi
while $O_{\mu\nu} (A,Z)$ can be written as:

$$O_{\mu\nu} (A,Z) = R_{\mu\nu} (f(A,Z))$$

\bi

\noi with

\bi

$$f(A,V) = \frac{1}{2} (1+ 4V)A+V; ~~~V= X,Y,Z .$$

\bi

Instead of working the field theory involving heavy baryons, we derive 
the Feynman rules for the effective theory by carrying out the $\frac{1}{m}$
expansion of the Feynman rules for BChPT and assume that the latter deffine
the effective Heavy Baryon Chiral Perturbation Theory [15]. Rather than 
repeating the procedure worked out by J-M, just recall that the spin 3/2 
fields of definite velocity are defined by:

$$\psi^\mu_v (x) = e^{-imv.x} \psi^\mu (x)$$

\bi

\noi where m stands for the nucleon mass in the chiral limit, {\it i.e.} the 
decuplet Lagrangian contains an explicit decuplet mass term proportional to 
$\Delta m = M -m$. 

\bi

In the heavy baryon limit $(m \to \infty)$, writing $P_\mu = m v_\mu + k_\mu$ 
and keeping leading terms in the 1/m expansion, the propagator (3) reduces to 
  
\begin{eqnarray}
i\Delta_{\mu\nu} (P,A)= \bigg\{ -g_{\mu\nu} &+& \frac{1}{3} \gamma_\mu \gamma_\nu - \frac{1}{3} (\gamma_\mu v_\nu - v_\mu \gamma_\nu ) \nonumber \\
&+& \frac{2}{3} v_\mu v_\nu \bigg\} \frac{P_+}{(v \cdot k-\Delta m + i \epsilon)} + {\cal O} \bigg( \frac{1}{m} ; A \bigg) \nonumber \\
&\equiv& i \Delta^0_{\mu\nu} + {\cal O} \bigg( \frac{1}{m} ; A \bigg) \nonumber
\end{eqnarray}

\noi where

$$P_+  = \frac{1+ \not v}{2}$$

\noi It is straigthforward to check that

\bi

\begin{equation}
\gamma^\mu \Delta^0_{\mu\nu} = \Delta^0_{\mu\nu} \gamma^\nu = 0
\end{equation}

\bi

Notice that relations (11) are reminiscent of the constraints (1) holding for 
the ~{\bf free} spin 3/2 fields. ~The important ~observation is ~that to 
~leading ~order in ~the ~1/m expansion, ~the spin 3/2 ~field ~behaves ---as 
~far as ~the ~constraints is concerned---like an on shell field; even if it is 
propagating. This implies that terms of the Lagragian containing the 
``off-shell" ~parameters X,Y,Z, give no contribution to the S matrix elements.
~Indeed, ~for ~any ~calculation regarding the Lagrangian (10), the vertices 
$O_{\mu\nu}$ and $O_{\mu\nu\alpha}$ connect to an external on-shell heavy 
baryon or to a heavy baryon propagator $\Delta^0_{\mu\nu}$. The constraint
for the free field (1), or relation (11) for the heavy propagator ensure in 
both cases the vanishing of the terms involving the ``off-shell" parameters.

\bi

These results suggests that to leading order in the 1/m expansion, the
constraints $\gamma^\mu \psi_\mu =0$ do not get modified by the interaction.
This lead us naturally to the question of the constraints in the interacting
theory.

\bi

The equation of motion for the spin 3/2 field derived from (10) is:

$$W_\mu \equiv \Lambda_{\mu\nu} \psi^\nu +i{\cal C} O_{\mu\nu} (A,X,Y) \psi 
\Delta^\nu +i {\cal H} O_{\mu\nu\alpha} \psi^\nu \Delta^\alpha =0$$

where
$$\Lambda_{\mu\nu} = P_\alpha \Gamma^\alpha_{\mu\nu} (A)-M ~B_{\mu\nu} (A)$$

The ~subsidiary ~conditions ~are ~derived ~by ~performing ~the ~following 
operations:
\begin{eqnarray}
\left[ \frac{1}{1+2A} \bigg( (1+A) \frac{\not P}{M}-1 \bigg) \gamma^\mu - 
\frac{2}{M} P^\mu \right] W_\mu &=& 0 \\
\frac{1+A}{1+2A} \gamma^\mu W_\mu -2 \gamma^{^0} W_{_0} &=& 0
\end{eqnarray}

These equations can be written as:

\bi
\begin{eqnarray}
&~&3(1+2A)M\gamma^\mu \psi_\mu =i{\cal C} \bigg(\gamma^\mu -\frac{2}{M} P^\mu
\bigg) O^\prime_{\mu\nu} (Z) \psi \Delta^\nu \nonumber \\
&-& i {\cal H} \{ \gamma^\mu - \frac{2}{M} P^\mu \} O^\prime_{\mu\beta} (X) 
\tau^\beta_{\nu\alpha} (A,Y) \psi^\nu \Delta^\alpha \\
\mbox {} \nonumber \\
&-&\frac{1}{2} \bigg(\frac{2}{3} \vec P \cdot \vec \gamma - M \bigg) (3(A+1) 
\gamma_{_0} \psi^0 -(3A+1) \vec \gamma \cdot \vec \psi)= P_i 
\psi^i_{3/2} \nonumber \\
&-&\frac{i {\cal C}}{2}\bigg((1+2Z) \gamma_\nu -2\gamma^{^0} g_{_{0\nu}}\bigg)
\psi \Delta^\nu - \frac{i {\cal H}}{2} \bigg((1+2X)\gamma_\beta -2 
\gamma^{^0}g_{_{0\beta}} \bigg) \nonumber \\
&~& \tau^\beta_{\nu\alpha} (A,Y) \psi^\nu \Delta^\alpha 
\end{eqnarray}

where
\begin{eqnarray}
O^\prime_{\mu\nu} (Z) &=& g_{\mu\nu} - \frac{1}{2} (1+2Z) \gamma_\mu 
\gamma_\nu \nonumber \\
\tau^\beta_{\nu\alpha} (A,Y) &=& g^\beta\,_\sigma \gamma_\alpha \gamma_5
R^\sigma\,_\nu (f(A,Y)) \\
\psi_{3/2}^i &=& (\delta^{i \ell} + \frac{1}{3} \gamma^i \gamma^\ell)
\psi_\ell \nonumber
\end{eqnarray}

Note that Eq. (14) implies

$$\gamma^\mu \psi_\mu =0 + {\cal O} \bigg( \frac{1}{m} , Z,X,Y \bigg)$$

\bi

\noi in agreement ~with ~our ~previous ~guess. ~Futhermore from Eq. (14) ~is 
~easily seen that ``off-shell" ambiguities appear in the next to leading order
of the 1/m expansion, hence ``off-shell" ambiguities free calculations can be 
carried on only to lowest order in the 1/m expansion of the Baryon Chiral 
Lagrangian. An implication of our argument is the need to keep track 
separately of the expansion parameter m (the baryon mass $\approx$ 1 GeV), and
$\Lambda_\chi$ ($\approx$ 1 GeV) even if they are of the same order of 
magnitude. $\Lambda_\chi$ is the parameter for the chiral expansion, on which 
there are no other limitation than usefullness, whereas consistency of the 
spin 3/2 theory restricts the 1/m expansion to the lowest order.

\bi

Summarizing, we pointed out that Heavy Baryon Chiral Perturbation Theory faces
the problems of the inconsistecies related to the interacting spin 
$\frac{3}{2}$ field theory [7,8,12,13]. We argued, by considering the 
$\frac{1}{m}$ expansion of the Feynman rules, that only the {\bf lowest} order
of the 1/m expansion is free of the so called ``off-shell" ambiguities. 
Furthermore, we have shown that the constraints --which assure the elimination
of the unphysical degrees of freedom contained in the Rarita-Schwinger 
spinor-- is spoiled by the interactions by terms of order $\bigg(\frac{1}{m} 
\bigg)$, implying thus consistency of $HBChPT$ only to lowest order of the 
expansion. We have also stressed the pertinence of dealing separately with the
chiral and the 1/m expansion.

\newpage

 \begin{center} {\large \bf REFERENCES}
\end{center}
\small
\begin{itemize}
\item[1.-] For a recent account see J.F. Donohgue, E. Golowich \& B.R. Holstein, ``Dynamics of the standard model" Cambridge Univ. Press (1992).
\item[2.-] E. Jenkins \& A. Manohar Phys. Lett. B255 (1991), 558.
\item[3.-] For a review see. B. Grinstein in ``Proceedings of the Workshop on High Energy Phenomenology" Eds. M.A. P\'erez y R. Huerta. World Scientific. (1992), 161.
\item[4.-] M.N. Butler {\it et.al.} Nucl. Phys. B399 (1993) 85. See also ref. 
2 
\item[5.-] J.W. Bos {\it et.al.} Phys. Rev. D51, (1995), 6308.
\item[6.-] W. Rarita \& J. Schwinger. Phys. Rev. 60, 61 (1941)
\item[7.-] C. Fronsdal Nuovo Cimento (suppl) IX, 416 (1958). See also J. Urias, Ph D. tesis,  Universite  Cat\'olique de  Louvain (1976), unpublished. 
\item[8.-] L.M. Nath, B. Etemadi \& J.D. Kimel, Phys. Rev. D3, 2153 (1971)
\item[9.-] S. Kamefuchi, L. O'Raifeartaigh \& A. Salam Nucl. Phys. 28, 529 (1961).
\item[10.-] M. El. Amiri, G. L\'opez \& J. Pestieau, Nucl. Phys. A 543 (1992) 673.
\item[11.-] R.D. Peccei, Phys. Rev. 181  1902 (1969), Phys. Rev. 176 1812 (1968)
\item[12.-] K. Jhonson \& E.C.G. Sudharsan, Ann. Phys. 13, 126, (1961).
	    G. Velo \& D. Zwanziger, Phys. Rev. 186, 1337 (1969).
\item[13.-] C.R. Hagen, Phys. Rev. D4, 2204 (1971).
\item[14.-] Davidson {\it et. al} Phys. Rev. D43, (1991), 71 \& M.G. Olsson and E.T. Ossypowski, Nucl. Phys. B87, 339 (1975), Phy. Rev. D17, 174 (1978).   
\item[15.-] M.B. Wise in ``New Symmetries in Strong Interactions". Lectures presented at the Lake Louise Winter Institute (1991).
\end{itemize} 
\end{document}